\documentclass[final,3p,twocolumn]{elsarticle}
\usepackage{graphics}
\pdfoutput=1

\usepackage{amssymb}

\journal{Journal of Alloys and Compounds}

\begin{document}

\begin{frontmatter}



\title{Transparent yttrium hydride thin films prepared by reactive sputtering}


\author[IFE]{T. Mongstad\corref{cor1}}
\author[Uppsala]{C. Platzer-Bj\"{o}rkman}
\author[IFE]{S. Zh. Karazhanov}
\author[IFE]{A. Holt}
\author[IFE]{J. P. Maehlen}
\author[IFE]{B. C. Hauback}

\address[IFE]{Institute for Energy Technology, P.O. Box 40, NO-2027 Kjeller, Norway}
\address[Uppsala]{Uppsala University, Box 534, SE-751 21 Uppsala, Sweden}

\cortext[cor1]{Corresponding author. Email: trygve.mongstad@ife.no Telephone: +47 99228200}

\begin{abstract}
Metal hydrides have earlier been suggested for utilization in solar cells. With this as a motivation we have prepared thin films of yttrium hydride by reactive magnetron sputter deposition. The resulting films are metallic for low partial pressure of hydrogen during the deposition, and black or yellow-transparent for higher partial pressure of hydrogen. Both metallic and semiconducting transparent YH$_{x}$ films have been prepared directly \emph{in-situ} without the need of capping layers and post-deposition hydrogenation. Optically the films are similar to what is found for YH$_{x}$ films prepared by other techniques, but the crystal structure of the transparent films differ from the well-known YH$_{3-\eta}$ phase, as they have an fcc lattice instead of hcp. 
\end{abstract}


\end{frontmatter}

\section{Introduction}
The present work is motivated by the possibility of using the semiconducting phase of yttrium hydride in a solar cell. Metal hydrides could be very interesting materials for application in solar cells because of the wide range of band gaps covered by the different semiconducting phases of hydrides \cite{Karazhanov2008}. This in combination with the abundance and non-toxicity of these materials make them candidate materials for use in future thin film solar cells and even for highly efficient multiple band gap third generation solar cells. 

Yttrium hydride is observed to form three different phases at room temperature and atmospheric pressure. The metallic solid solution $\alpha$-phase, the metallic dihydride $\beta$-phase (YH$_{2}$)  and the insulating trihydride $\gamma$-phase (YH$_{3}$) \cite{Vadja1995}. An optical transition is observed during the transformation from the metallic $\beta$-phase to the transparent, insulating $\gamma$-phase \cite{Huiberts1996}. Studies show that $\gamma$-phase yttrium hydride is semiconducting with a band gap of 2.63 eV \cite{Gogh2001}, corresponding to a wavelength of 470 nm. Thus it could be used in a solar cell for violet to ultraviolet light or as a top cell in a multiple band gap solar cell stack. A third possibility is the usage of yttrium hydride for antireflection coating for silicon solar cells \cite{Karazhanov2010a}.

Earlier, yttrium hydride films were prepared by hydrogenation of metallic films in a slow process that could take several days. Using a palladium cap layer dramatically shortened the process time \cite{Huiberts1996a}. However, a palladium cap layer is not always desirable, firstly because the layer is light absorbing. Secondly the Pd cap layer method would introduce additional manufacturing steps that would not comply with the high levels of efficiency that are necessary in the production of solar cells. Thirdly, Palladium is also an expensive element, which is not desirable to use in low-cost solar cells. A fourth issue is the stress that will result from the hydrogenation of metallic films, as a 19.5\% volume increase results of the hydrogenation of pure metallic yttrium to the semiconducting $\gamma$-phase of yttrium hydride \cite{Gogh2001}. The stress from this volume expansion could deteriorate the material quality and constitute a problem for utilization in electronic devices such as solar cells.

Deposition of metallic yttrium dihydride films by the use of pulsed laser deposition (PLD) \cite{Dam2003} and by evaporation of Y under low pressure of H$_{2}$ \cite{Hayoz2000} has been reported earlier, but we have not found reports of direct \emph{in-situ} deposition of insulating and transparent films of $\gamma$-phase yttrium hydride. 

\section{Experimental methods}

Yttrium hydride (YH$_{x}$) films were deposited using pulsed DC magnetron sputtering in a Leybold Optics A550V7 in-line sputtering system. A metallic yttrium target (99.99\%) was used with argon (purity: 5N) and hydrogen (6N) gas in the chamber during deposition. The chamber was pumped down to 10$^{-6}$ mbar between the depositions. The depositions were done at pressures from $2\times10^{-3}$ to $1\times10^{-2}$ mbar. The DC power of the sputter was varied from 100 to 1000 W and pulsed at a frequency of 70 kHz with a reverse cycle of 4 $\mu$s. Pulsed DC magnetron sputtering (PDC) is used to avoid arcing and building up charge on the surface of the target, as the target can become insulating for high hydrogen injection levels \cite{Berg2005}. The hydrogen to argon flow rate was varied from pure argon to a 1:1 ratio between the hydrogen and argon flow rates. Films were deposited on glass (26 mm $\times$ 76 mm) and quartz (Suprasil, 10 mm $\times$ 20 mm) substrates. The thickness of the deposited films were from 100 nm to 2.6 $\mu$m.

We did \emph{ex-situ} characterization of the electrical, optical, structural and compositional properties of the samples. The thickness of the samples was determined by using profilometry. Electrical resistivity was determined by a four-point probe measurement setup. The optical transmission and reflection of the films were measured by an Ocean Optics spectrometer setup with a halogen lamp supported by a deuterium lamp for increased intensity in the ultraviolet region. Reflection measurements were done with an integrating sphere to collect both diffuse and specularly reflected light. 

Structural characterization was done by X-ray diffraction (XRD) using a Bruker-Siemens D5000 diffractometer with Cu K$\alpha$ radiation in $\theta$-2$\theta$ , parallel beam geometry. The diffraction data were analysed by the Rietveld whole-profile refinement method \cite{Rietveld1969} using the General Structure Analysis System (GSAS) software \cite{Larson1994}. The peak shapes were described using pseudo-Voigt-type function. Preferred orientation was modelled by the March-Dollase method \cite{Dollase1986}.

\section{Results and discussion}

\begin{figure}[t]%
\centering
\includegraphics[width=7cm]{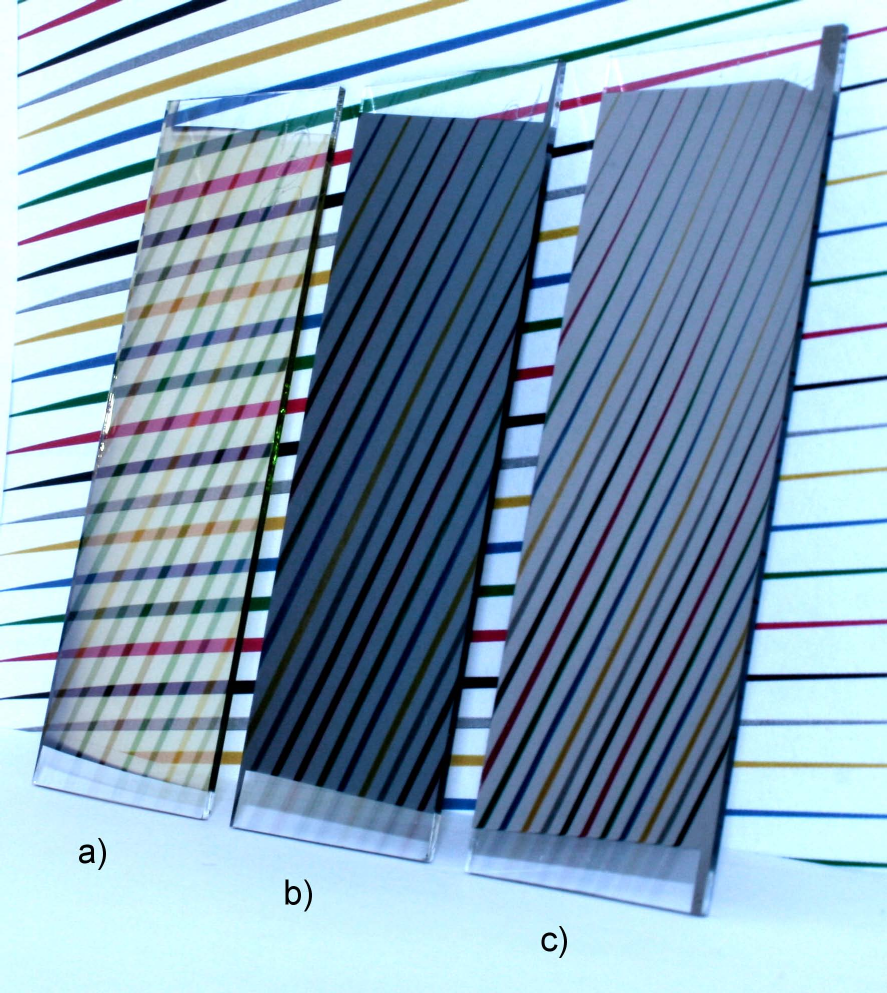}%
\caption{A photograph of yttrium hydride films deposited on glass. Vertical lines visible on the samples visualize the reflection, whereas the horizontal lines on the background visualize the transmission of the samples. a) A 440 nm thick transparent yttrium hydride film. The appearance of this sample resembles that of the $\gamma$-phase, but as explained in the text, the crystal structure differs from the normal $\gamma$-phase. The film has high transmission of light, but appears yellowish due to the absorption of blue and green light with photon energies above the band gap.  b) A 380 nm thick black yttrium hydride film deposited with reactive hydrogen in the process. This sample is mainly composed of the $\beta$-phase of yttrium hydride. c) A 370 nm thick yttrium metal film deposited without hydrogen.}%
\label{photo}%
\end{figure}

Films of metallic yttrium, metallic yttrium hydride and insulating yttrium hydride were deposited by varying the hydrogen flow during deposition. For zero or low hydrogen flow, metallic $\alpha$-phase yttrium films were obtained. For higher hydrogen flow ratios, black or transparent films were formed. Figure \ref{photo} shows the visual appearance of the three different classes of films obtained from the experiments. The black films were conductive with a resistivity in the order of 10$^{-3}$ $\Omega$cm, approximately one order of magnitude higher than our measurements for metallic yttrium films. The transparent films were found to be highly insulating, with resistivity in the order of 10$^{4}$ $\Omega$cm.

\subsection{Optical properties}

\begin{figure}[t]
\centering
\includegraphics[width=7cm]{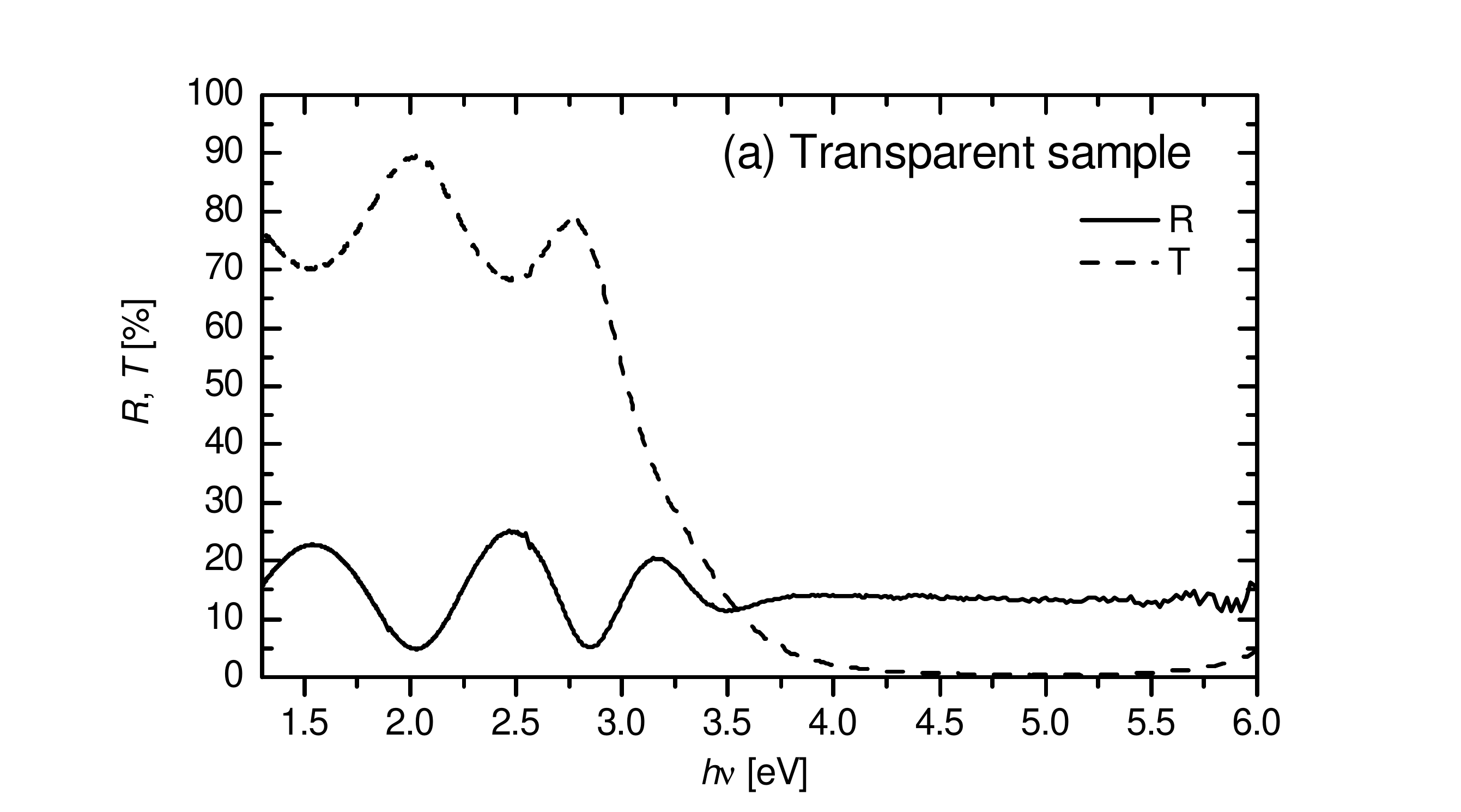}\\
\includegraphics[width=7cm]{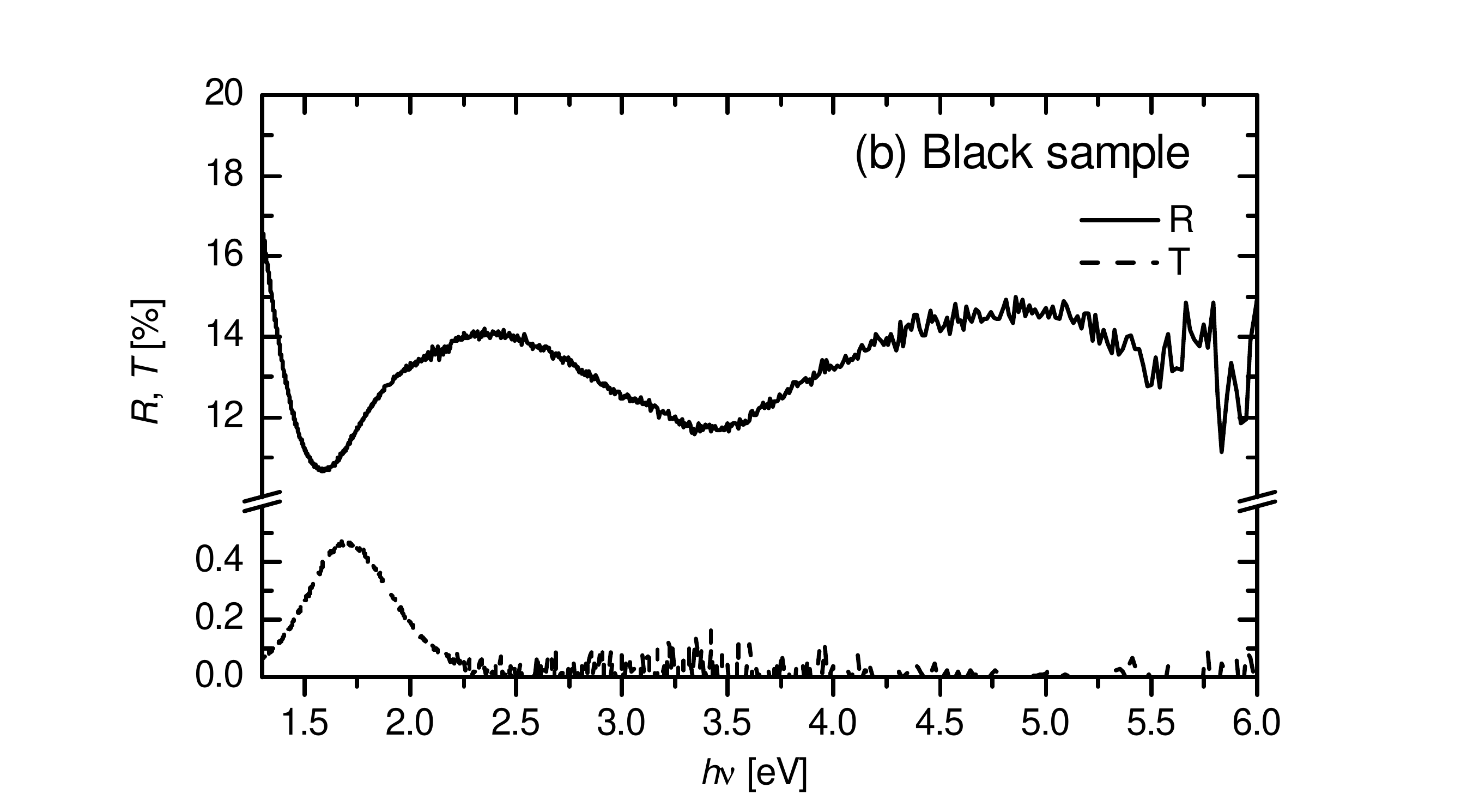}\\
\includegraphics[width=7cm]{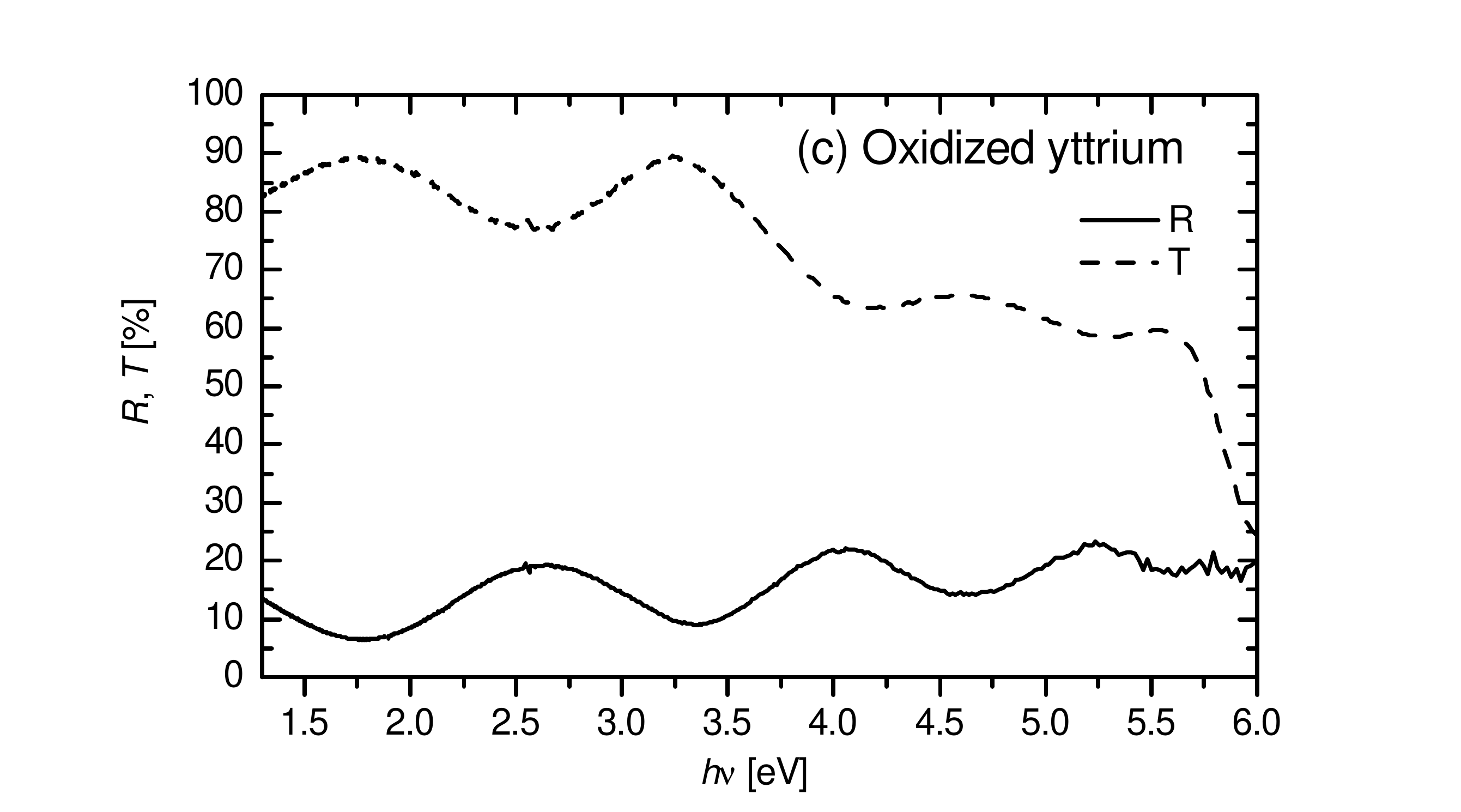}%
\caption{Reflection and transmission for films deposited on quartz substrates. a) For a 280 nm thick transparent yttrium hydride film. b) Reflection and transmission for a 360 nm thick black yttrium hydride film. Note the difference in scale for b). For comparison we here also present the reflection and transmission spectra of c) an 165 nm thick oxidized yttrium film.}%
\label{RT}%
\end{figure}

The reflection and transmission for one of the transparent samples deposited on a quartz substrate is displayed in Figure \ref{RT}a. The data shows an absorption edge between 2.5 and 3 eV. For visible light with photon energy $h\nu < 2.5$ eV (wavelength $\lambda > 500$ nm), the films are highly transparent. The low absorption is exceptional for yttrium hydride thin films, Pd-covered YH$_3$ films with comparable thicknesses are normally observed to absorb up to 80\% of the light even for photon energies smaller than the band gap \cite{Gogh2001, Lee1999, Lokhorst2003}. The oscillations in the reflection and transmission are due to interference effects, and can be modelled with a film with a thickness 280 nm and a refractive index $n \approx 2.3$.

The absorption coefficient can be estimated from the reflection and transmission data from the approximation $\alpha = d^{-1} \ln [(1-R)/T]$, where $d$ is the thickness of the film, and $R$ and $T$ are the measured reflection and transmission, respectively. Using this formula we find that the present films have a high absorption coefficient in the order of 10$^{5}$ cm$^{-1}$ for photon energies above the band gap, which is close to what is found for materials that are utilized in thin film solar cells \cite{Bube1998}. With such a high absorption a thin layer of one to two micrometer of absorber material is sufficient to absorb most of the photons with energy higher than the band gap in a solar cell.

The optical band gap can be found by extrapolating the linear region of the plotted values of $(\alpha h \nu)^{m}$ as a function of the photon energy $h \nu$. $m = 1/2$ for indirect band gap transitions or amorphous materials, and $m = 2$ for direct band gap transitions. The extrapolated line crosses the $h \nu$-axis at the band gap energy $E_{g}$ \cite{Tauc1974}. For the data presented in Figure \ref{RT}a, we found an indirect optical band gap of 2.6 eV. All transparent samples had band gap values in the range 2.2-2.7 eV. This is in accordance with the band gap $E_{g} = 2.63$ eV van Gogh et al. \cite{Gogh2001} found for an epitaxial $\gamma$-phase yttrium hydride. They also mentioned that the band gap of polycrystalline films varied in the range 2.3-2.7 eV. Fitting our data for a direct band gap gives a value approximately 0.5 eV higher than for the indirect band gap.

Figure 2b shows the reflection and transmission data obtained for a black film. The reflectivity is low, around 10-15\%, over the entire measured spectrum. The minima in reflection at 1.6 eV and 3.5 eV are not interference fringes, but can be attributed to the inherent reflectivity of yttrium dihydride. The same minima were found for the reflection of light on thin yttrium dihydride films prepared by hydrogenation of yttrium, in the work of Sakai et al. \cite{Sakai2004}. The transmission is very low, but a transmission window with a peak value at 1.7 eV lets up to 0.5\% of the light through. This transmission window is also characteristic for the metallic $\beta$-phase of yttrium hydride \cite{Gogh2001, Sakai2004}.

For reference, the reflection and transmission spectra of an oxidized yttrium sample are also presented in Figure \ref{RT}c. This is to show that the strong absorption above 2.6 eV seen in Figure \ref{RT}a is because of the hydride and is not related to the substrate or oxidized regions of the deposited yttrium hydride film. The oxidized yttrium was prepared by annealing a metallic yttrium film  at 850 $^{\circ}$C for 2 minutes in air. The thickness of the sputter deposited yttrium film was 145 nm before, and 165 nm after annealing. Notice the absorption edge at 5.6 eV, close to the band gap of Y$_{2}$O$_{3}$.

We have not been able to quantify the hydrogen content of our samples, but comparation of the optical measurements to similar studies \cite{Gogh2001,Sakai2004,Enache2005} gives a strong indication that the transparent samples have a composition close to YH$_{3}$ and the black samples have a composition close to YH$_{2}$.

\subsection{Structural properties}

The crystal structures of the three phases of yttrium hydride are well known from the literature \cite{Vadja1995}; the $\alpha$-phase (yttrium metal/solid solution) crystallizes with a hcp structure, the $\beta$-phase (YH$_{2}$) with fcc and the transparent $\gamma$-phase (YH$_{3}$) with a hcp structure. However, we found that the structures of the transparent and the black films were almost indistinguishable. They were both found to form an fcc structure, with a small shift in the lattice parameter. The metallic yttrium films deposited without hydrogen were found to be in the expected hcp structure.

We were surprised to find the transparent film with the fcc structure, but it was not completely unexpected. Firstly, the very similar hydride of lanthanum shows an optical transition from metallic to transparent upon hydrogenation which is not accompanied by any structural transition from fcc to hcp lattice \cite{Huiberts1996,Gogh2001}. Secondly, a hydrogen-induced optical transition without change in crystal structure has also been observed for hydrogenated nanocrystalline PLD-grown yttrium dihydride films \cite{Dam2003}. A third similar finding was reported by Van der Molen et al. \cite{Molen2001}, who found that transparent yttrium hydride films were stabilized in the fcc structure by the addition of magnesium. Their coevaporated Y$_{z}$Mg$_{1-z}$ films maintained the fcc structure after switching from metallic to transparent when the Mg content was higher than z = 0.1. They found the optical properties to be similar for fcc and hcp films. In addition, a structural transition for YH$_{3}$ from hcp to fcc under high pressures (several GPa) has also been predicted \cite{Ahuja1997} and observed for bulk YH$_{3}$ samples \cite{Palasyuk2005}.

\begin{figure}%
\centering
\includegraphics[width=7cm]{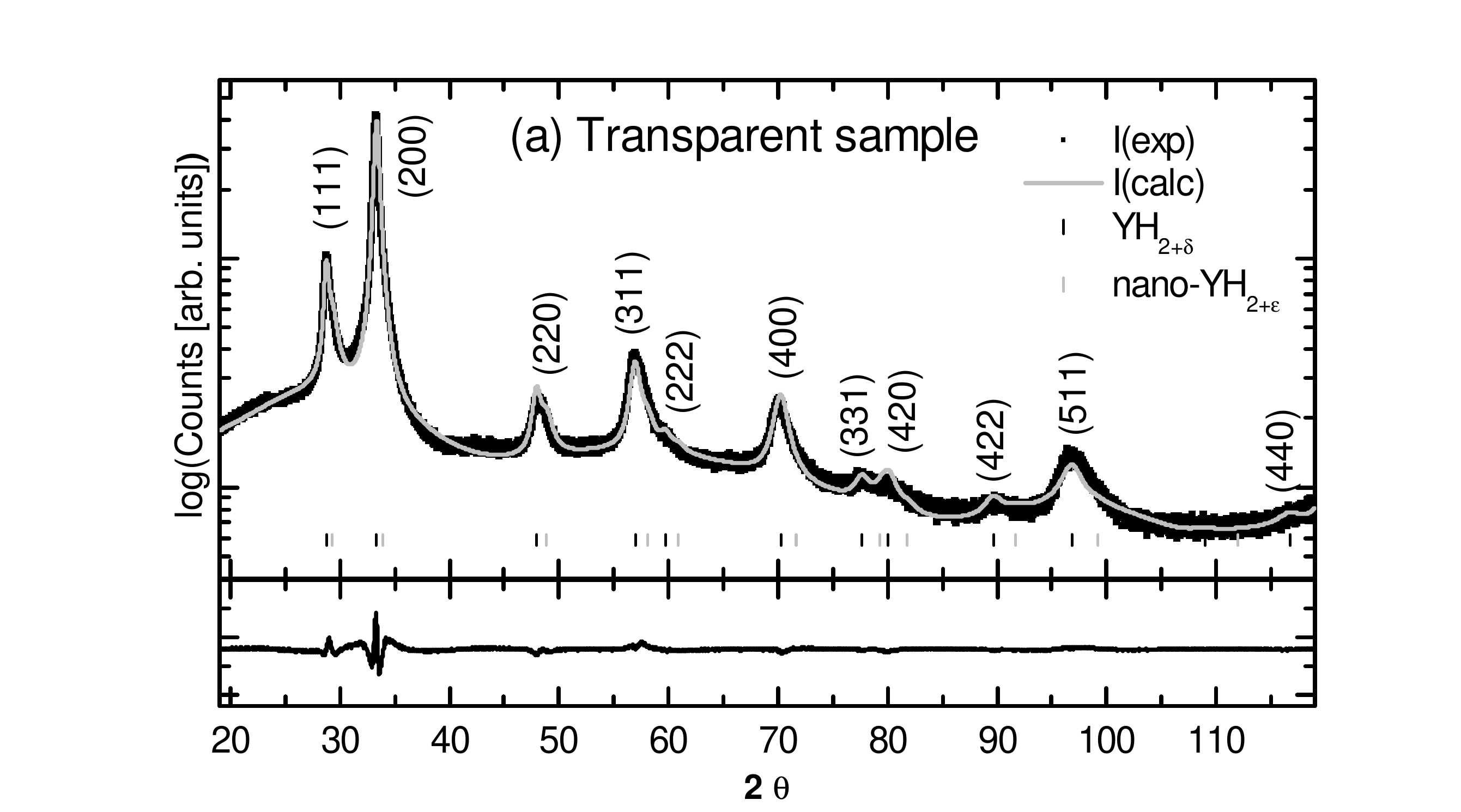}\\
\includegraphics[width=7cm]{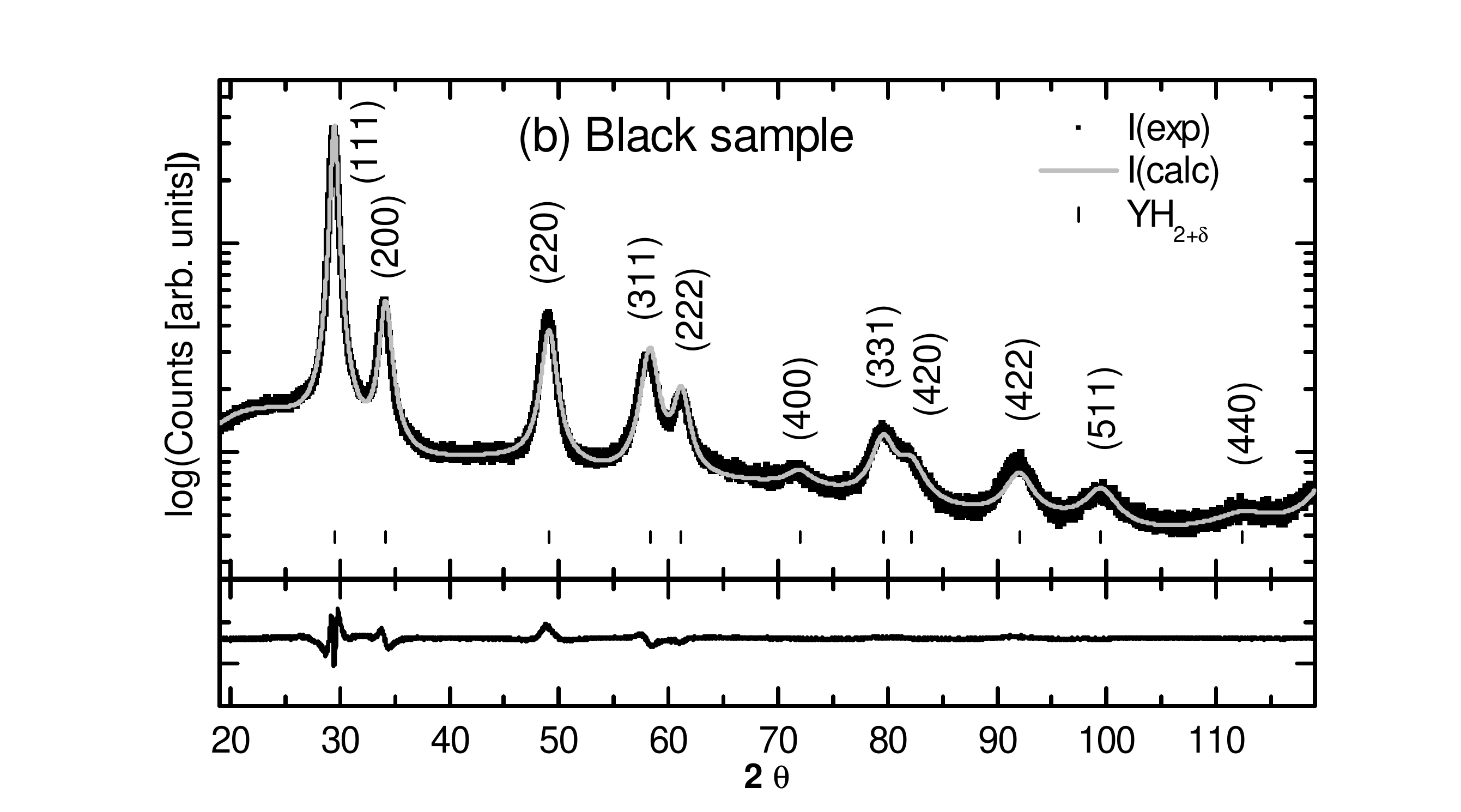}\\
\includegraphics[width=7cm]{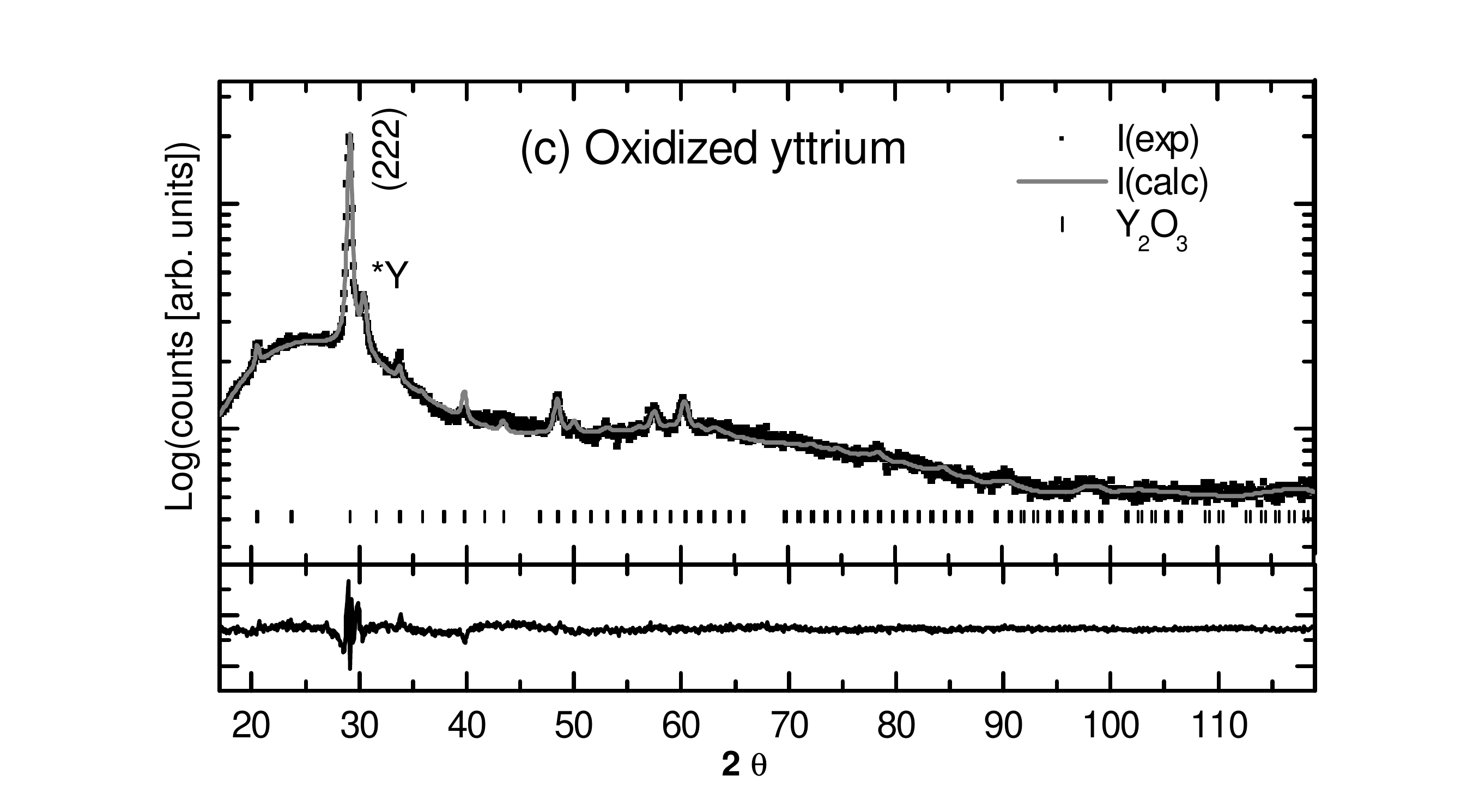}%
\caption{Rietveld-type fits and XRD patterns collected at room temperature ($\lambda = 1.5406$ \AA) showing observed (squares) and calculated (line) values in the upper panes (log scale) and the difference between the observed and calculated values in the lower panes (linear scale). The positions of the Bragg reflections are shown as ticks. a) For a transparent sample (space group $Fm\overline{3}m$, $a = 5.346(2)$ \AA, $R_{wp} = 0.726$). b) For a black sample (space group $Fm\overline{3}m$, $a = 5.262(2)$ \AA, $R_{wp} = 0.0745$). For comparison we here also present the XRD spectra of c) an oxidized yttrium film (space group $Ia\overline{3}$, $a = 10.64(1)$ \AA, $R_{wp} = 0.0484$). An yttrium 002-peak at 30.5$^\circ$ is also visible. The thicknesses of the films from which these XRD data were obtained, were 2.6 $\mu$m, 2.8 $\mu$m and 240 nm, respectively.}%
\label{XRD}%
\end{figure}

Figures \ref{XRD}a and \ref{XRD}b display XRD patterns and Rietveld-type whole-profile refinements for a transparent and a black sample deposited on a glass substrate. The crystal planes from which the Bragg peaks originate is indicated on the figure, according to an $Fm\overline{3}m$ space group with the corresponding lattice parameter. For comparison, the XRD pattern of an Y$_2$O$_3$ film is also shown in Figure 3c. There is a noticeable difference in the intensity distribution over the peaks for the two hydride samples in Figures 3a and 3b. A reasonable explanation of this is that there is more orientational growth in the $\left\langle 100\right\rangle$-direction for the transparent sample. Rietveld-type whole-profile fitting of the XRD data was performed where the preferred orientation was described using March-Dollase models choosing the (111) plane as preferred orientation plane for the black sample and the (100) plane for the transparent sample. To describe the asymmetry in the peaks observed for the transparent sample, a secondary cubic YH$_2$ phase with size-broadened peaks (nano-sized grains) was added to the refinements. The residua $R_{wp}$ of all the presented fits were less than 8\%.   

The diffraction patterns of the transparent and the black yttrium hydride films correspond to a fcc structure with lattice parameters $a_{transp} = 5.35$ {\AA} and $a_{black} = 5.26$ \AA. These values were representative for all our transparent and black samples. This corresponds to molar volume of 23.05 cm$^3$ for the transparent sample and 21.9 cm$^3$ for the black sample.

If we assume that the composition of the transparent samples is close to YH$_{3}$, the expanded unit cell in comparison with the black sample (YH$_{2}$) is in agreement with the findings of van der Molen et al. \cite{Molen2001}, who found that the fcc unit cell expanded from  $a = 5.20$ {\AA} to 5.28 {\AA} when going from fcc YH$_{2}$ to fcc YH$_{3}$. Tkacz et al. found the same unit cell parameter $a = 5.28$ {\AA} for a high-pressure fcc phase of YH$_{3}$ at 8 GPa. In the work of van Gogh et al. \cite{Gogh2001} on La$_{1-z}$Y$_{z}$H$_{x}$ one may perform an extrapolation of the data on the fcc films to obtain a lattice parameter of $a = 5.33$ {\AA} for stoichiometric YH$_{3}$ films in the fcc structure \cite{Molen2001}. 

As mentioned earlier, the optical transmission and reflection of the black samples resemble what is found for yttrium dihydride. For stoichiometric YH$_{2}$ ($\beta$-phase), $a = 5.20$ {\AA} is found for both bulk samples at room temperature \cite{Vadja1995} and thin films \cite{Gogh2001}. The larger unit cell parameter we found for the black samples indicate that the composition differ from stoichiometric YH$_{2}$. Both differences in hydrogen content and the presence of oxygen may influence the unit cell parameter.

\subsection{Oxygen content in samples}

The films prepared in this study were not capped by Pd, as usually done when yttrium hydride films are prepared by metal deposition and post process hydrogenation. Therefore, we have no surface protection against oxidation. Yttrium forms a stable oxide, and is known to dissociate water or oxygen from air to form yttrium oxide (Y$_{2}$O$_{3}$). The present samples were kept in air and no visible degradation was observed, even after several months of storage. Some surface oxidation should be expected, but it is likely that this oxidation will affect a very thin layer, as the oxide itself forms a barrier for further oxidation \cite{Curzon1978}. 

It is difficult to distinguish Y$_{2}$O$_{3}$ from the $\beta$-phase of yttrium hydride by XRD. Yttrium oxide forms a bcc lattice with a lattice parameter of 10.60 Å$ = 2 \times 5.30$ {\AA} \cite{Xu1997}, and can therefore give a XRD pattern that is similar to the XRD measurements for the black and transparent samples (Fig. \ref{XRD}). However, the optical behavior of Y$_{2}$O$_{3}$ is very different from what we observed (Fig. \ref{RT}), as Y$_{2}$O$_{3}$ is transparent with an optical band gap of 5.5 eV \cite{Xu1997}. It is therefore probable that transparent yttrium hydride is responsible for the observed optical properties.

Films with capping layers of yttrium, aluminum and molybdenum were also prepared. The capping layers were deposited by DC magnetron sputtering, on top of the hydride films without any air exposure in between the depositions. Similar structural results were obtained for these films, but the optically transparent state was only seen with the Mo capping. For yttrium, the reason for not obtaining the transparent film could be that hydrogen easily diffuses into the yttrium cap layer, and leaves the sample in the metallic dihydride state. The aluminium cap layer reacted with the underlying yttrium hydride film, and is not suitable as capping layer for this kind of sample. The Mo cap layer did not react with the sample, and from samples deposited on glass we could observe the transparent sample below a 150 nm thick capping layer of molybdenum. For molybdenum, formation of hydride is known to happen only at pressures of several kbar \cite{Driessen1990}, so we assume that the Mo layer does not absorb any hydrogen from the underlying film. 

It is also probable that oxygen can be incorporated in the films during the deposition process. An exact determination of the oxygen and hydrogen content of the samples is not a straightforward task, and we are currently working on a quantification of the actual oxygen and hydrogen concentration in the samples. A preliminary analysis show that the oxygen content in the samples is between 5 and 30 at\%. A full analysis of the oxygen and hydrogen content will be the subject of a subsequent paper from our group.

\subsection{Yttrium hydride as a solar cell material}
As to the suggested application of metal hydrides in solar cells, some sort of internal electric field has to be set up in a semiconductor to separate the generated charge carriers. This is normally done by setting up a p-n junction by introducing different impurities in a single semiconducting material (homojuction), or by depositing different semiconducting materials with different doping in a stack (heterojunction) \cite{Bube1998}. In either case a careful study of the electric properties of semiconducting yttrium hydride needs to be done before proceeding to make more complex device structures implementing yttrium hydride. Some work on electrical properties has been done \cite{Huiberts1996a,Enache2005}, but it is uncertain if these results are also valid for our \emph{in-situ} prepared films that exhibit a cubic crystal structure. 

\section{Conclusion}

Films of metallic yttrium, metallic black yttrium hydride and insulating transparent yttrium hydride were deposited \emph{in-situ} by using PDC sputtering of a metallic yttrium target in an atmosphere of argon and hydrogen. This method differs from other methods for preparation of yttrium hydride films in that it is a direct deposition technique that requires no post deposition hydrogenation step and that no palladium cap is needed. This could be advantageous when optical and electrical parameters are important. The insulating, or semiconducting, films were highly transparent for light with photon energy below the band gap. We found an indirect optical band gap of 2.2-2.7 eV for our samples. 

The transparent samples are observed to have a cubic fcc structure, which is interesting from a fundamental point of view, as high quality transparent films of cubic yttrium hydride at ambient pressure have earlier only been achieved by stabilization of the films by the addition of magnesium. The level of oxidation of the films is not determined, but experiments with capping of the deposited films with molybdenum show that the transparent state is also achievable for films that have not been exposed to air.

\section*{Acknowledgements}

Funding from the Norwegian Research Council through the NANOMAT program is acknowledged. We also thank professor R. Griessen, Netherlands, for fruitful discussions of our results.





\bibliographystyle{elsarticle-num}
\bibliography{bibliography}







\end{document}